\documentclass[english,aps,prd,twocolumn]{revtex4-1}
\usepackage[T1]{fontenc}
\usepackage[latin9]{inputenc}
\usepackage{amsmath}
\usepackage{amssymb}
\usepackage{graphicx}

\makeatletter
\@ifundefined{textcolor}{}
{%
 \definecolor{BLACK}{gray}{0}
 \definecolor{WHITE}{gray}{1}
 \definecolor{RED}{rgb}{1,0,0}
 \definecolor{GREEN}{rgb}{0,1,0}
 \definecolor{BLUE}{rgb}{0,0,1}
 \definecolor{CYAN}{cmyk}{1,0,0,0}
 \definecolor{MAGENTA}{cmyk}{0,1,0,0}
 \definecolor{YELLOW}{cmyk}{0,0,1,0}
}

\makeatother

\usepackage{babel}
\begin{document}

\title{Origin of Symmetric PMNS and CKM Matrices }

\author{Werner Rodejohann$^a$ and Xun-Jie Xu$^{a,b}$ }

\affiliation{$^a$Max-Planck-Institut f$\ddot u$r Kernphysik, Postfach 103980, D-69029
Heidelberg, Germany\\
$^b$Institute of Modern Physics and Center for High Energy Physics,
Tsinghua University, Beijing 100084, China\\
(\,werner.rodejohann@mpi-hd.mpg.de and xunjie.xu@gmail.com\,)
}
\begin{abstract}
The PMNS and CKM matrices are phenomenologically close to symmetric, and a symmetric form could be used as zeroth-order approximation for both matrices. We study the possible  theoretical origin of this feature in flavor symmetry models. 
We identify necessary geometric properties of discrete flavor symmetry groups that can lead to symmetric mixing matrices. Those properties are actually very 
common in discrete groups such as $A_{4}$, $S_{4}$ or $\Delta(96)$. 
As an application of our theorem, we generate a symmetric lepton mixing scheme 
with $\theta_{12}=\theta_{23}=36.21^{\circ};\theta_{13}=12.20^{\circ}$ and $\delta=0$, 
realized with the group $\Delta(96)$. 
\end{abstract}
\maketitle

\section{\label{intro}Introduction}
The properties of the fermion mixing matrices are expected to give important hints on the 
underlying flavor physics. Flavor symmetries \cite{Altarelli:2010gt} are an attractive and most often studied approach to explain the rather different structure of the 
Pontecorvo-Maki-Nakagawa-Sakata (PMNS) and Cabibbo-Kobayashi-Maskawa (CKM) %\cite{CKM1,CKM2} 
mixing matrices. 
%\cite{PMNS1,PMNS2,PMNS3}. 
Literally hundreds of models have been proposed in the literature, applying many possible discrete groups in order to explain lepton and quark mixing. Instead of adding simply another model to that list, we study in this paper an interesting possible property of both the CKM and PMNS matrices. Namely, despite the fact that the CKM mixing is a small while the PMNS mixing is large, both can to reasonable precision be estimated to be symmetric. 
The symmetric form of the CKM matrix has early been noticed and studied in many references \cite{Branco:1990zq,Samal:1991im,Kielanowski:1989wc,Rosner:1989yn,Samal:1991sf,Tanimoto:1992kw,Gonzalez:1997hb,
Xing:1997jg,Chaturvedi:2003hg}.
 After neutrino oscillation was well established the possible symmetric PMNS matrix 
also attracted some attention \cite{wn,wn2,zee,lm,sm,Xing:2002sx,Liao:2014vla}.
%  those matrices has been noticed and studied in many references 
The symmetric form discussed in these references includes the manifestly symmetric case ($U=U^{T}$) 
and the Hermitian case ($U=U^{\dagger}$). 
It is easy to get the relation 
\begin{equation}
(U=U{}^{T})\Rightarrow(|U|=|U|^{T})\Leftarrow(U=U^{\dagger})\label{eq:0110}
\end{equation}
by taking absolute values, 
which implies any physical prediction from $|U|=|U|^{T}$ can also
be used in the other two cases $U=U{}^{T}$ or $U=U^{\dagger}$.
Both of them are special cases of $|U|=|U|^{T}$, which is what we mean by symmetric mixing matrix from now on. 

Using the global fits of the  CKM \cite{PDG} and PMNS \cite{NeuFitLisi} matrices, one finds:  
\begin{equation}
|U_{{\scriptscriptstyle \textrm{CKM}}}|=\left(\begin{array}{ccc}
\left(\begin{array}{c}
0.97441\\
0.97413
\end{array}\right) & \left(\begin{array}{c}
0.22597\\
0.22475
\end{array}\right) & \left(\begin{array}{c}
0.00370\\
0.00340
\end{array}\right)\\
\left(\begin{array}{c}
0.22583\\
0.22461
\end{array}\right) & \left(\begin{array}{c}
0.97358\\
0.97328
\end{array}\right) & \left(\begin{array}{c}
0.0426\\
0.0402
\end{array}\right)\\
\left(\begin{array}{c}
0.00919\\
0.00854
\end{array}\right) & \left(\begin{array}{c}
0.0416\\
0.0393
\end{array}\right) & \left(\begin{array}{c}
0.99919\\
0.99909
\end{array}\right)
\end{array}\right)\label{eq:1208-1}
\end{equation}
\begin{equation}
|U_{{\scriptscriptstyle \textrm{PMNS}}}|=
\left(\begin{array}{ccc}
\left(\begin{array}{c}
0.845\\
0.791
\end{array}\right) & \left(\begin{array}{c}
0.592\\
0.512
\end{array}\right) & \left(\begin{array}{c}
0.172\\
0.133
\end{array}\right)\\
\left(\begin{array}{c}
0.521\\
0.254
\end{array}\right) & \left(\begin{array}{c}
0.698\\
0.455
\end{array}\right) & \left(\begin{array}{c}
0.782\\
0.604
\end{array}\right)\\
\left(\begin{array}{c}
0.521\\
0.254
\end{array}\right) & \left(\begin{array}{c}
0.698\\
0.455
\end{array}\right) & \left(\begin{array}{c}
0.782\\
0.604
\end{array}\right)
\end{array}\right).\label{eq:1208-5}
\end{equation}
Here the upper (lower) values in each entry are upper (lower) bounds
of the  matrix elements. The CKM matrix has been measured to a
high precision (here we show the $1\sigma$ range) and the relations
$|U_{12}|=|U_{21}|$, $|U_{23}|=|U_{32}|$ are still well compatible with data.  
The relation $|U_{13}|=|U_{31}|$ is, however, not fulfilled by data. 
As a symmetric mixing matrix requires that \cite{Branco:1990zq,wn} 
\begin{equation}\label{eq:ut}
|U_{31}|^2 - |U_{13}|^2 = |U_{12}|^2 - |U_{21}|^2 = |U_{23}|^2 - |U_{32}|^2=0 \,,
\end{equation}
we have an interesting option, namely, that some flavor symmetry or other mechanism 
generates $|U_{12}|=|U_{21}|$, $|U_{23}|=|U_{32}|$ but $U_{13} =  U_{31} = 0$. Higher order corrections, which are frequently responsible for the smallest mixing angles, are then the source of non-zero $|U_{13}| \neq  |U_{31}|$, as well as for CP violation. Rather trivially, matrices with only one mixing angle are symmetric, the same holds for the unit matrix.

%Actually due to the smallness
%of $|U|_{13}$ and $|U|_{31}$, they are very unstable if the other
%elements change a little while the unitarity still holds. For example,
%if we take $|U|_{11}=0.97420$ and $|U|_{21}=0.225522$, then from
%the unitarity we have $|U|_{31}=\sqrt{1-|U|_{11}^{2}-|U|_{21}^{2}}=0.0086$.
%Now if we change $|U|_{11}$ to $|U|_{11}=0.974233$ we will get $0.0031$
%for $|U|_{31}$. So the mismatch of $|U|_{13}$ and $|U|_{31}$ is
%not a seriously problem in discussing symmetric CKM mixing. 
%If there
%is a very tiny correction to the yukawa coupling of quarks in the
%Standard Model, then $|U|_{31}$ and $|U|_{13}$ will change to other
%values not far from zero. The correction may come from RG running
%or the underlying flavor symmetry which has been studied in many references{[}{]}.

The symmetry conjecture for the PMNS mixing is less compatible with data, as shown by the 3$\sigma$ bounds in Eq.\ (\ref{eq:1208-5}) \footnote{
some elements have non-Gaussian errors so actually we compute in Eq.\ (\ref{eq:1208-5}) the $99.7\,\%$ CL bounds}. 
Similar to the quark sector, the 13- and 31-elements are incompatible with symmetry (the other two relations between the elements are also not favored by data), and a similar situation as mentioned above for the CKM matrix might be realized. Of course, one could also imagine that an originally symmetric mixing matrix is modified by higher order corrections, VEV-misalignment, RG-effects or other mechanisms that have been studied in the literature.  

For completeness, we give the phenomenological prediction of a symmetric mixing matrix, using the standard parametrization of the CKM and PMNS mixing matrices \cite{wn}: 
\begin{equation}\label{eq:hr}
|U_{13}| = \frac{\sin \theta_{12} \, \sin\theta_{23}}
{\sqrt{1 - \sin^2 \delta \, \cos^2 \theta_{12}\, \cos^2 \theta_{23}} + \cos \delta\,\cos\theta_{12}\,\cos\theta_{23}}
\end{equation}
This is the unique physical prediction of both $|U|=|U|^{T}$ and
$U=U{}^{T}$. Note that $|U|=|U|^{T}$ has only one prediction as
the unitarity requires that the relation in Eq.\ (\ref{eq:ut}) is fulfilled, so once we set
$|U_{13}|^{2}=|U_{31}|^{2}$ we immediately get $|U|=|U|^{T}$. It
is also the unique prediction of $U=U{}^{T}$ because any 3-by-3 unitary $U$ with 
$|U|=|U|^{T}$ can be transformed to a new $U'$ satisfying $U'=U'{}^{T}$ 
simply by rephasing \cite{Branco:1990zq,wn}. Note that this does not hold for more than 3 generations. 
The Hermitian case $U=U^{\dagger}$ has not only the prediction of 
Eq.\ (\ref{eq:hr}) but also CP conservation \cite{zee}. Thus, it would predict 
$\sin \theta_{13} = \tan \theta_{12} \, \tan \theta_{23}$.

Despite the apparent deviation from their symmetric forms, one can still use 
it as an attractive zeroth-order Ansatz and attempt to study its theoretical origin. 
One option, put forward in \cite{sm}, is that a single unitary matrix 
$V$ diagonalizes all mass matrices of quarks 
and leptons at leading order, in addition to the $SU(5)$ relation 
$m_d = m_\ell^T$ between the down quark and charged lepton mass matrices. 
While it is difficult to embed this in realistic mass spectra in GUTs, the predictions of this scenario are that  $U_{{\scriptscriptstyle \textrm{CKM}}}=V^{\dagger}V=1$ 
while $U_{{\scriptscriptstyle \textrm{PMNS}}}=V^{T}V$ is symmetric.

In this paper, we study the the origin of symmetric mixing matrices from an underlying
flavor symmetry. We prove a theorem which links geometric properties of discrete symmetry 
groups to the symmetric form of the mixing matrices. This theorem, explained in detail in 
Section \ref{sec:A-Theorem-for}, holds only for subgroups of $SO(3)$ with real representations, and can interestingly be realized in the most often studied groups $A_4$ and $S_4$. 
We find a modification that holds in the complex case in Section \ref{sec:Generalization}, that could be applied to subgroups of $SU(3)$, for instance to $\Delta(96)$. We use this to reproduce a previously studied, and actually symmetric, mixing scenario for the PMNS matrix in Section \ref{sec:Application}.

Since our analysis links the properties of the symmetry group with the mixing matrix, we end our introduction with a summary on how the generators of the group can be related to the matrices diagonalizing the mass matrices, following the strategy developed in  \cite{Lam:2008rs,Lam:2008sh,Lam:2011ag}. 
 In general, if a flavor symmetry group $G$ is applied to, for instance, the 
lepton sector, then it must be broken to two residual symmetries $G_{\ell}$ and $G_{\nu}$ acting 
 on the charged lepton sector and neutrino sector: 
\begin{equation}
G\rightarrow\begin{cases}
G_{\nu}: & S^{T} M_{\nu} S=M_{\nu}\\
G_{\ell}: & T^{\dagger} M_{\ell} T=M_{\ell} \,.
\end{cases}\label{eq:1210}
\end{equation}
Here the left-handed neutrino (assumed to be Majorana) mass matrix
$M_{\nu}$ is invariant under the transformation $S^{T} M_{\nu} S$
for $S\in G_{\nu}$ and $M_{\ell}$ (defined by $m_{\ell} m_{\ell}^{\dagger}$, where $m_{\ell}$ is the charged lepton mass matrix) is the effective mass matrix of
left-handed charged leptons, invariant under $T^{\dagger} M_{\ell} T$.
Then the diagonalizing matrices $U_{\nu}$ and $U_{\ell}$ defined by
\begin{eqnarray}
M_{\nu} & = & U_{\nu} D_{\nu} U_{\nu}^{T};\label{eq:1210-2}\\
M_{\ell} & = & U_{\ell} D_{\ell} U_{\ell}^{\dagger};\label{eq:1210-1}
\end{eqnarray}
can be directly determined by $S$ and $T$ according to \cite{Lam:2008rs,Lam:2008sh,Lam:2011ag}
\begin{eqnarray}
U_{\nu}^{\dagger} S U_{\nu} & = & D_{S}\,;\label{eq:1210-3}\\
U_{\ell}^{\dagger} T U_{\ell} & = & D_{T}\label{eq:1210-4}\,,
\end{eqnarray}
where all $D$ are diagonal matrices.
Note that $U_{\nu}$ obtained from Eq.\ (\ref{eq:1210-3}) does not include 
Majorana phases which rephase each column of 
$U_{\nu}$. 
%to make the $D_{\nu}$ in Eq.\ (\ref{eq:1210-2}) real and diagonal. 
Eq.\ (\ref{eq:1210-3}) is independent of such rephasing which 
means the Majorana phases are not determined by flavor symmetries 
in this approach. 
For Majorana
neutrinos $G_{\nu}$ has to be a direct product of two $\mathbb{Z}_{2}$,  
i.e.\ $\mathbb{Z}_{2}\otimes\mathbb{Z}_{2}$ to fully determine the
mixing in neutrino sector\cite{He:2011kn}. For quarks, we have the same framework
but since they are Dirac fermions, we do not have to be limited to 
$\mathbb{Z}_{2}\otimes\mathbb{Z}_{2}$. In this case note that 
$S^T$ and $U_{\nu}^T$ in Eqs.\ 
(\ref{eq:1210},\ref{eq:1210-2}) should be replaced with 
$S^{\dagger}$ and $U_{\nu}^{\dagger}$.

\section{\label{sec:A-Theorem-for}A Theorem for $|U|=|U|^{T}$}
In this paper, as mentioned above, we define symmetric mixing matrix as
\begin{equation}
|U|=|U|^{T}\label{eq:1203}
\end{equation}
rather than the original definition of $U=U^{T}$ or the Hermitian case $U=U^{\dagger}$.
 The phenomenology is the same for $|U|=|U|^{T}$ and $U=U^{T}$, but more general than the Hermitian case.

Before we formulate our theorem that links geometrical properties of the flavor symmetry group to a symmetric mixing matrix, we will first define the geometric concepts which
will be used. 

\begin{figure}
\centering

\includegraphics[width=8.1cm]{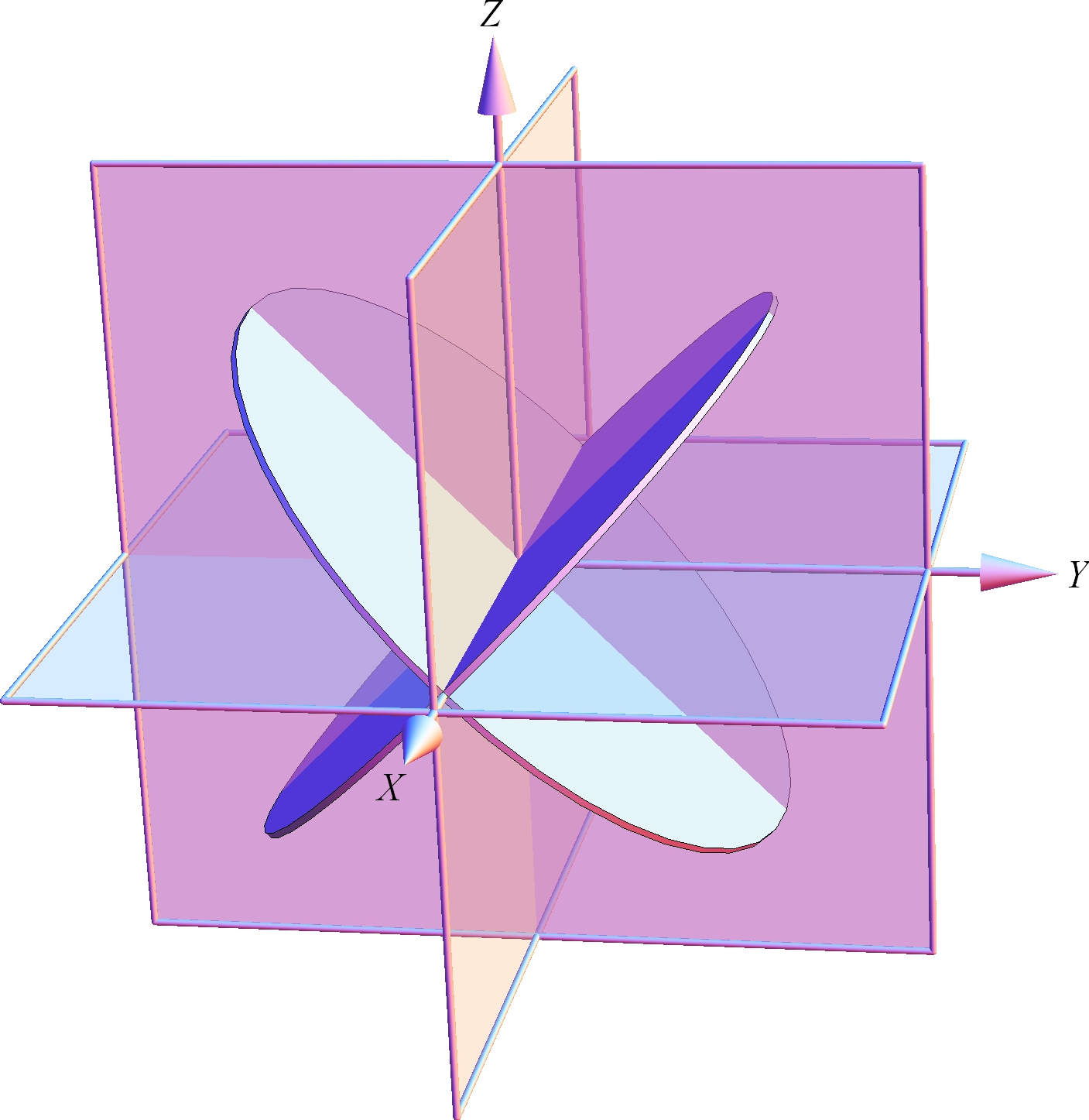}

\protect\caption{\label{fig:2-bisect}The geometrical relation of the mirror planes
and the bisecting planes. The mirrors are placed on $y-z$, $z-x$
or $x-y$ planes. The two round disks are called bisecting planes because
they bisect all the square mirror planes. The bisecting planes are
boundaries of octants. }
\end{figure}

\begin{figure}
\centering

\includegraphics[width=8.1cm]{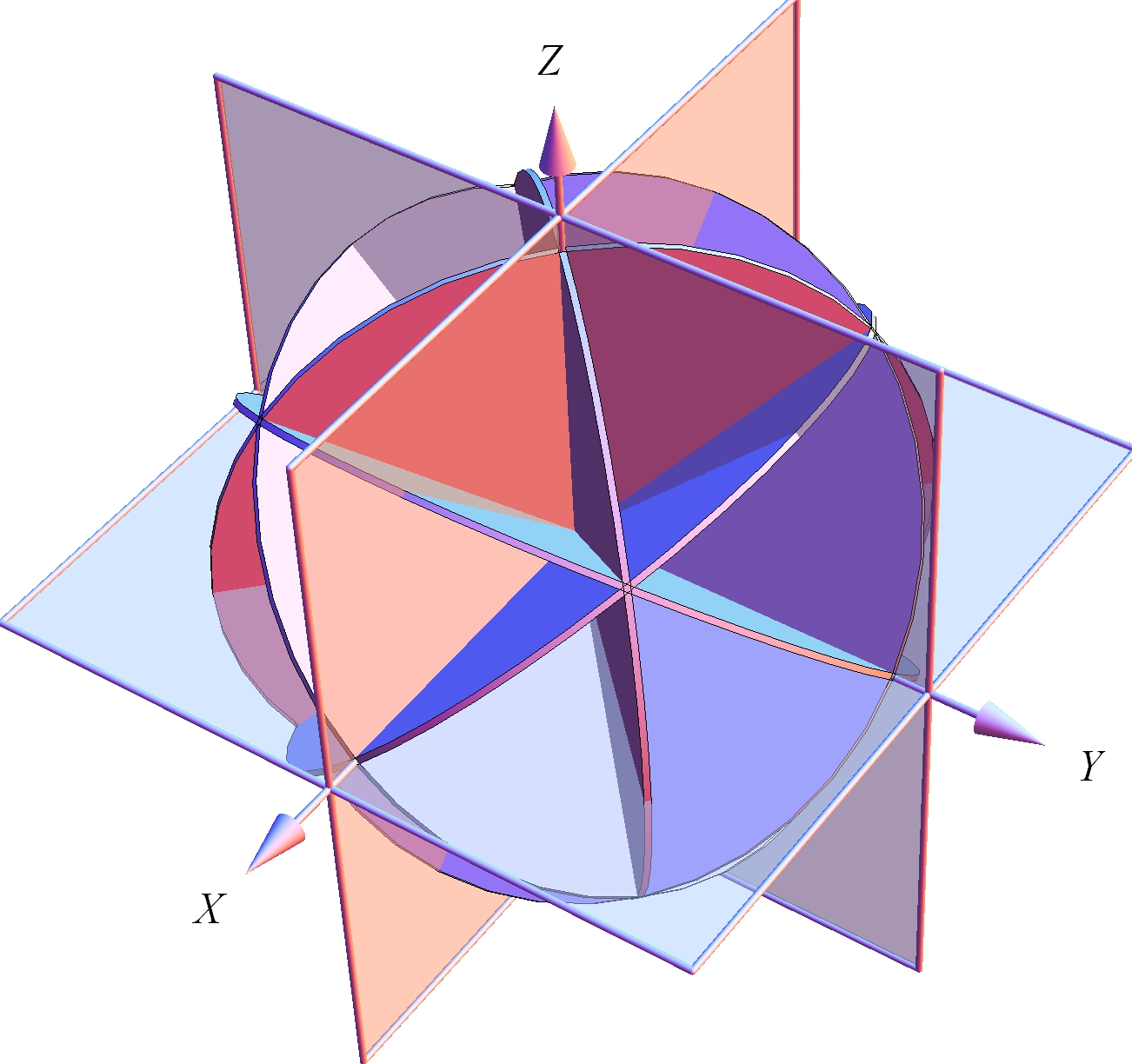}

\protect\caption{\label{fig:6-special}The complete collection of all six possible
bisecting planes and their geometrical relation with the mirror planes.
A rotational symmetry with its axis on one of these bisecting planes can give,  
according to Theorem A, a symmetric mixing $|U|=|U|^{T}$.}
\end{figure}

The $\mathbb{Z}_{2}$ symmetries used in neutrino sector are actually
just reflections or $180^{\circ}$ rotations (the difference between
them is trivial, any $180^{\circ}$ rotation in 3-dimensional space
can be changed to a reflection if we add an overall minus sign
and vice versa). In 3-dimensional flavor space, going w.l.o.g.\ in the diagonal neutrino 
basis, the $\mathbb{Z}_{2}$ transformations correspond to putting minus
signs to neutrino mass eigenstates: $\nu_{i}\rightarrow-\nu_{i}$. 
The combined $\mathbb{Z}_{2}\otimes\mathbb{Z}_{2}$ in 3-dimensional flavor 
space corresponds to two reflections with respect to the direction of neutrino
mass eigenstates. Or, in a picture we are more familiar with, if there 
are planes whose normal vectors are neutrino mass eigenstates,
the $\mathbb{Z}_{2}$ symmetries are just the mirror symmetries of those planes. 
Since $\mathbb{Z}_{2}\otimes\mathbb{Z}_{2}$ contains commutative
mirror transformations or since the mass eigenstates are orthogonal,
the mirrors should be perpendicular to each other, as shown in Fig.\ \ref{fig:2-bisect}
by translucent squares. 
If the transformation from $G_{\ell}$ 
is real (we will discuss the complex case later) in flavor space,
it is an $SO(3)$ transformation which can always be represented by a
rotation. 
If the rotation axis is on the \emph{bisecting} plane, which is 
defined as the plane that bisects the two mirror squares, or as the boundaries
of octants in the diagonal neutrino basis, then we define that $G_{\ell}$
\emph{bisects} the two $\mathbb{Z}_{2}\otimes\mathbb{Z}_{2}$. We
show two bisecting planes in Fig.\ \ref{fig:2-bisect}, 
while all bisecting planes are shown in Fig.\ \ref{fig:6-special}. 
This gives now all definitions necessary for our theorem.\\

{\bf Theorem A:}

{\it If an $SO(3)$ subgroup $G$ contains two non-commutative Abelian subgroups $G_{\nu}$
and $G_{\ell}$, and if $G_{\nu}$ is isomorphic to 
$\mathbb{Z}_{2}\otimes\mathbb{Z}_{2}$ while $G_{\ell}$
\emph{bisects} the $\mathbb{Z}_{2}\otimes\mathbb{Z}_{2}$, then $G$ 
as a flavor symmetry can produce a symmetric mixing matrix. }\\

The definition of \emph{bisection} and symmetric mixing are given
previously. The subgroups $G_{\nu}$ and $G_{\ell}$ are required to be Abelian
because the residual flavor symmetries are always Abelian \cite{Lam:2008rs,Lam:2008sh,Lam:2011ag},
and non-commutative so that the mixing is non-trivial. 

The proof of this theorem will be obvious after we introduce the general
$SO(3)$ rotation and the diagonalization below (see Eqs.\ (\ref{eq:1129})
and (\ref{eq:1130})). The $\mathbb{Z}_{2}$ symmetries are applied to Majorana
neutrinos and the bisecting rotation to charged leptons. 
However, one can also apply the theorem to quarks and obtain a symmetric 
CKM mixing. In the case of Dirac fermions, $\mathbb{Z}_{2}$ is not necessary
but sufficient. We will comment further on CKM mixing later.  
Because the axis of a bisecting rotation can be rotated on its bisecting
planes, there are infinite bisecting rotations. Hence, Theorem A can produce 
infinite symmetric mixing matrices with one degree of freedom. Note
that the unitary matrix with the constraint $|U|=|U|^{T}$ has only one prediction,  
see Eq.\ (\ref{eq:hr}). 
% The geometrical relation of the $Z_{2}$ planes or axes (i.e.\ the
%square planes or $x$, $y$ and $z$ axes) with the \emph{bisecting
%planes} determines a symmetric PMNS mixing $|U|=|U|^{T}$. 

\begin{figure}
\centering

\includegraphics[width=8.1cm]{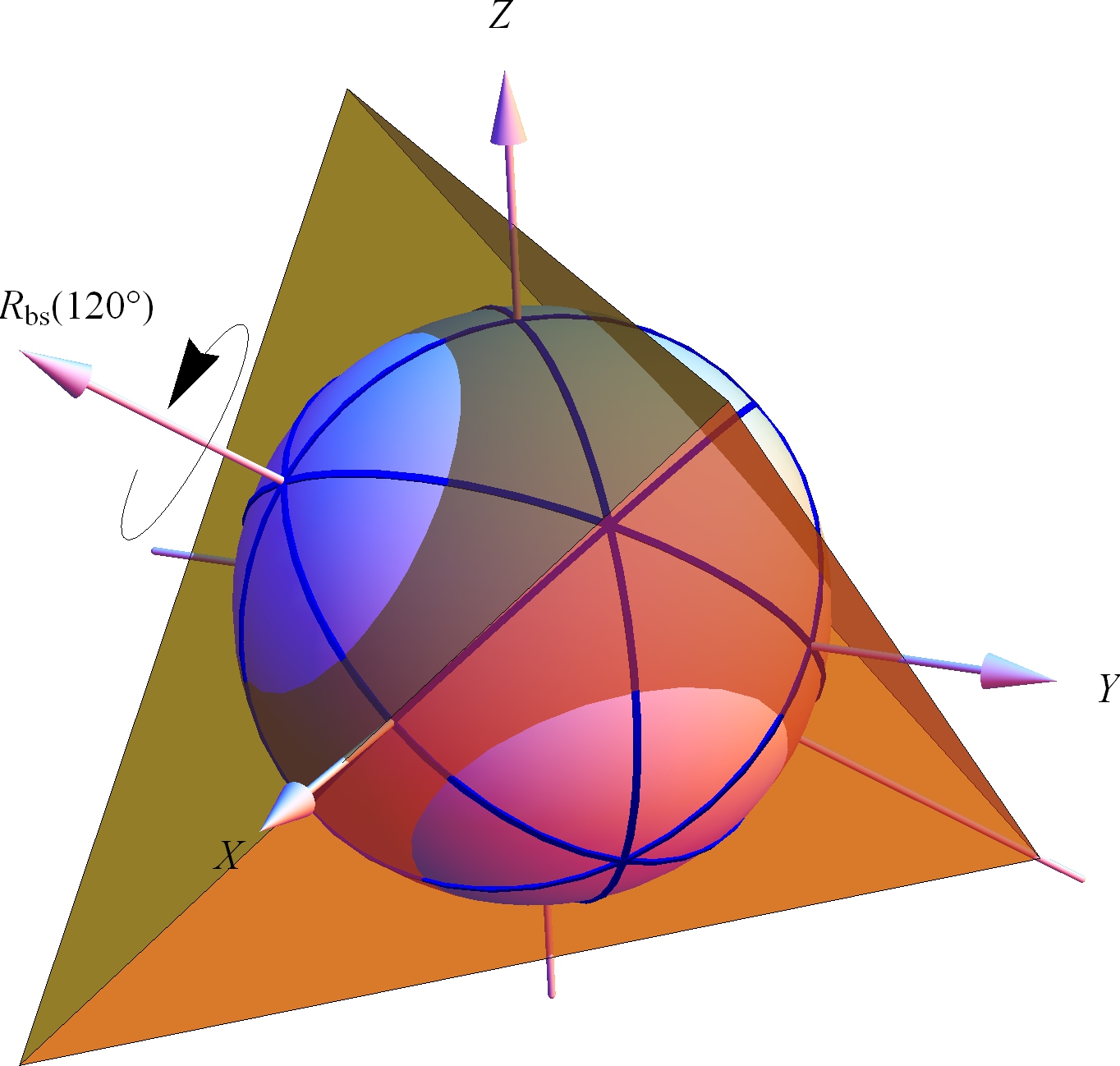}

\protect\caption{\label{fig:Tetrahedron}Tetrahedron symmetry. The dark 
blue circles show the bisecting planes. The axes of the $120^{\circ}$ rotational symmetries  
of the tetrahedron are on those planes, therefore according to Theorem A the 
tetrahedral group as a flavor symmetry can produce a symmetric
mixing matrix $|U|=|U|^{T}$. }
\end{figure}

Actually a lot of discrete flavor symmetries satisfy the conditions
required by Theorem A, for example the tetrahedral group $T$ and
octahedral group $O$ which are just the widely used $A_{4}$ and $S_{4}$
flavor symmetry groups, respectively (for geometrical interpretations on
$A_{4}$ and $S_{4}$ see, for e.g. \cite{He:2012yt}). 
We can see in  Fig.\ \ref{fig:Tetrahedron} that if we choose the three $180^{\circ}$ rotational
axes as $x$, $y$ and $z$ axes, which penetrate the tetrahedron
through the two central points of two edges, then the bisecting planes
are determined, as shown by dark blues circles. The tetrahedron is
also invariant under the $120^{\circ}$ rotation marked in Fig.\ \ref{fig:Tetrahedron},
which is a bisecting rotation since the axis is on three bisecting
planes. In explicit formulae, we say the tetrahedron is invariant
under the following rotations,
\begin{equation}
R_{\rm bs}=\left(\begin{array}{ccc}
0 & -1 & 0\\
0 & 0 & -1\\
1 & 0 & 0
\end{array}\right);\label{eq:0108}
\end{equation}
\begin{equation}
S_{1}={\rm diag}(1,-1,-1)\,;\thinspace S_{2}={\rm diag}(-1,1,-1)\,.\label{eq:0108-1}
\end{equation}
Here $R_{\rm bs}$ is the $120^{\circ}$ bisecting rotation and $S_{i}(i=1,2)$
are the $180^{\circ}$ rotations around the $x$ and $y$ axes. Eq.\ (\ref{eq:0108})
can be obtained by requiring that $R_{\rm bs}(1,0,0)^{T}=(0,0,1)^{T}$, which 
means $R_{\rm bs}$ rotates the $x$ axis to the $z$ axis, as well as 
the other two relations $R_{\rm bs}(0,1,0)^{T}=(-1,0,0)^{T}$ 
and $R_{\rm bs}(0,0,1)^{T}=(0,-1,0)^{T}$. 

If $R_{\rm bs}$ and $S_{i}(i=1,2)$ are the residual symmetries of the
charged lepton sector and neutrino sector respectively, i.e.\ 
\begin{equation}
R_{\rm bs}^{\dagger} M_{\ell} R_{\rm bs}=M_{\ell}\,;\,\,S_{i}^{T} M_{\nu} S_{i}=M_{\nu}\,,\label{eq:0108-2}
\end{equation}
then according to Eqs.\ (\ref{eq:1210-3},\ref{eq:1210-4}) we can compute $U_{\ell}$
and $U_{\nu}$ from $R_{\rm bs}$ and $S_{i}$. The result is 
\begin{eqnarray}
U_{\ell} & = & \frac{1}{\sqrt{3}}\left(\begin{array}{ccc}
1 & \omega & \omega^{2}\\
-1 & -\omega^{2} & -\omega\\
1 & 1 & 1
\end{array}\right);\thinspace U_{\nu}=1\,.\label{eq:0108-3}
\end{eqnarray}
We see that $U_{\ell}$ is the Wolfenstein matrix, up to trivial signs. 
Therefore in this case the PMNS matrix $U=U_{\ell}^{\dagger} U_{\nu}$
is symmetric. i.e.\ $|U|=|U|^{T}$. 

\begin{figure}
\centering

\includegraphics[width=8.1cm]{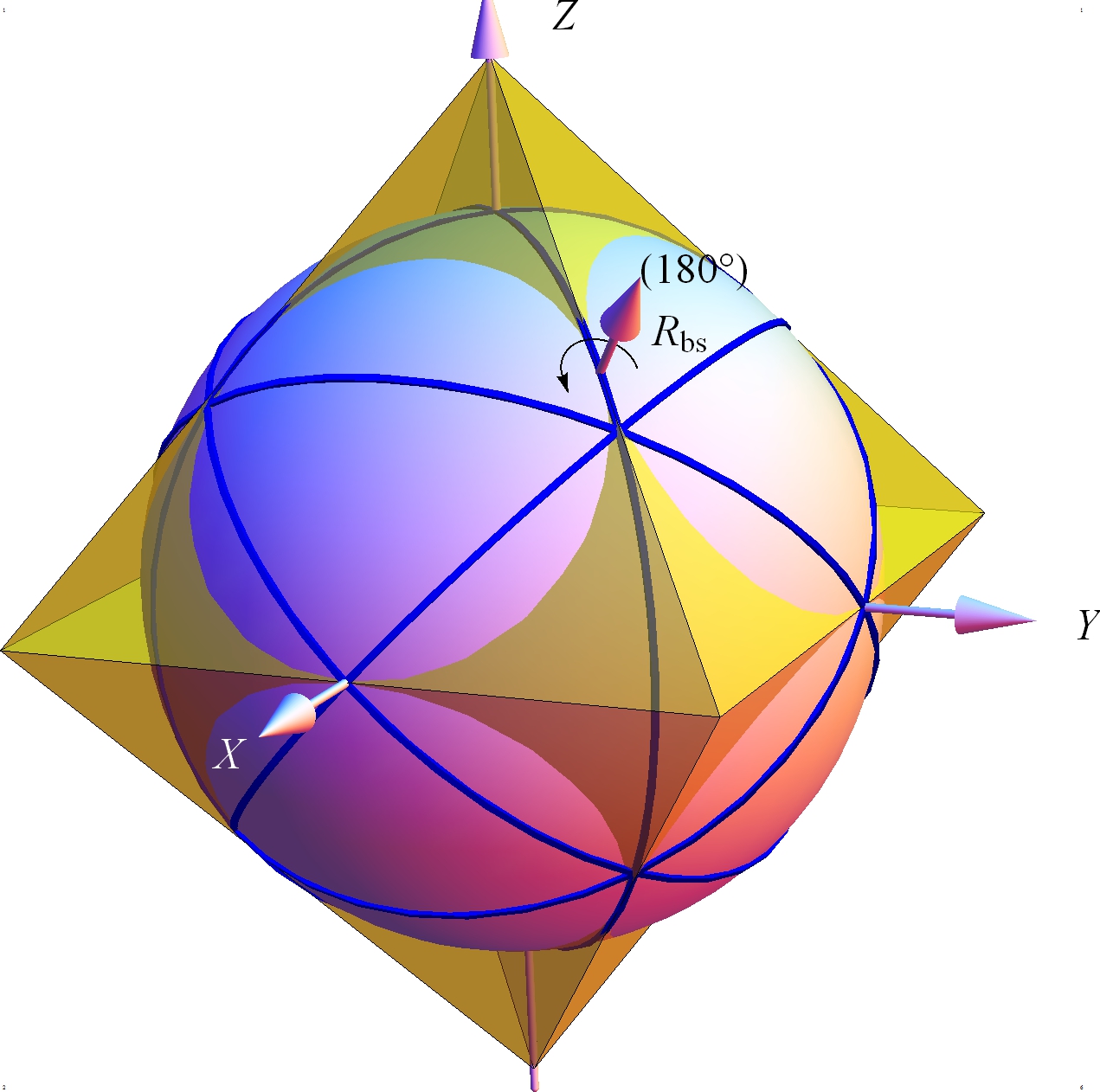}

\protect\caption{\label{fig:Octahedron}Octahedron symmetry. Similar to Fig.\ \ref{fig:Tetrahedron}, according to Theorem A the 
octahedral group can also be used to produce a symmetric mixing matrix  $|U|=|U|^{T}$.
}
\end{figure}

As another example we show in Fig.\ \ref{fig:Octahedron} that the
octahedral symmetry, which is isomorphic to the widely used $S_{4}$
symmetry, also has the required properties for Theorem A. 
Figs.\ \ref{fig:Tetrahedron} and \ref{fig:Octahedron} show that the properties 
required by Theorem A are quite common in discrete groups with 3-dimensional
irreducible real representations. \\

Now we present the theorem in explicit formulae. 
For simplicity, we choose a basis under which the mirror has a normal
vector $(1,0,0)^{T}$, $(0,1,0)^{T}$ or $(0,0,1)^{T}$ so the mirror
symmetry is just a reflection with respect to the $y-z$, $z-x$ or $x-y$
planes. Then the mirror transformations through $y-z$, and $z-x$
planes are
\[
S_{1}={\rm diag}(-1,1,1) \thinspace\mbox{ and } \thinspace S_{2}={\rm diag}(1,-1,1)\,
\]
respectively. Under this basis, the normal vectors $\boldsymbol{n}=(n_{1},n_{2},n_{3})^{T}$
of the six bisecting planes satisfy one of the six conditions:
\begin{equation}
|n_{i}|=|n_{j}|\,,\thinspace(i,j=1,2,3;\thinspace i\neq j)\,,\label{eq:1202-1}
\end{equation}
i.e.\ $n_{1}=\pm n_{2}$, $n_{2}=\pm n_{3}$ or $n_{1}=\pm n_{3}$.
The bisecting rotations $R_{\rm bs}$ with such an axis have special forms
which we will show below. The neutrino and charged
lepton mass matrices are now invariant under transformations of ($S_{1}$,
$S_{2}$) and $R_{\rm bs}$ respectively: 
\begin{eqnarray}
S_{i}^{T} M_{\nu} S_{i} & = & M_{\nu}\,;\thinspace(i=1,2)\label{eq:1201}\\
R_{\rm bs}^{\dagger}M_{\ell}R_{\rm bs} & = & M_{\ell}\,.\label{eq:1203-1}
\end{eqnarray}
Diagonalizing these matrices with the transformation
\begin{equation}
M_{\nu}=U_{\nu}D_{\nu}U_{\nu}^{T}\,,\thinspace\,\, M_{\ell}=U_{\ell}D_{\ell}U_{\ell}^{\dagger}\label{eq:1202}
\end{equation}
gives the PMNS mixing matrix $U=U_{\ell}^{\dagger}U_{\nu}$. According to 
Theorem A, $U$ will be symmetric with proper ordering of the eigenvectors.

Since we choose the basis in which $S_{1}$ and $S_{2}$ are diagonal,
$M_{\nu}$ is constrained to be diagonal by Eq.\ (\ref{eq:1201}), hence 
$U_{\nu}$ is diagonal. As discussed at the end of the Introduction, diagonalization of $(S_{1}, S_{2})$
and $R_{\rm bs}$ will give $U_{\nu}$ and $U_{\ell}$. So actually
the key point of Theorem A can be stated as follows:

\emph{For any $SO(3)$ matrix $R$, if $R=R_{\rm bs}$ there must
be an unitary matrix $U$ which is symmetric ($|U|=|U|^{T}$) and
can diagonalize $R$. The converse is also true which
means $R=R_{\rm bs}$ is the necessary and 
sufficient condition for $|U|=|U|^{T}$.}

Thus, in the basis we choose, we have a bisecting rotation to generate 
$|U_{\ell}|=|U_{\ell}|^{T}$ and the mirror symmetries to make $U_{\nu}$
diagonal, therefore we get a symmetric PMNS matrix. In the above discussion 
we have explained the theorem in a specific basis, however, the physical
result is independent of any basis. One can choose another basis where
the mirrors are not on the $x-y$, $y-z$ and $z-x$ planes, in which case the neutrino
sector is not diagonal and in general $|U_{\ell}|\neq|U_{\ell}|^{T}$. 
However, the geometrical relation of the bisecting planes and the mirror
planes makes sure that the product $U_{\ell}^{\dagger}U_{\nu}$ is
symmetric.

As for the explicit form of the bisecting rotation $R_{\rm bs}$, we should
first introduce the general rotation. The most general
rotation in Euclidean space which rotates the whole space around an
axis $\boldsymbol{n}=(n_{1},n_{2},n_{3})^{T}$ ($\boldsymbol{n}. \boldsymbol{n}=1$)
by an angle $\phi$ is
\begin{widetext}
\begin{equation}
R(\boldsymbol{n},\phi)=\left(\begin{array}{ccc}
n_{1}^{2}+c\left(n_{2}^{2}+n_{3}^{2}\right) & (1-c)n_{1}n_{2}+sn_{3} & -sn_{2}+(1-c)n_{1}n_{3}\\
(1-c)n_{1}n_{2}-sn_{3} & c+n_{2}^{2}-cn_{2}^{2} & sn_{1}+(1-c)n_{2}n_{3}\\
sn_{2}+(1-c)n_{1}n_{3} & -sn_{1}+(1-c)n_{2}n_{3} & c+n_{3}^{2}-cn_{3}^{2}
\end{array}\right),\label{eq:1129}
\end{equation}
where $c=\cos\phi$ and $s=\sin\phi$. 
\end{widetext}

One can check that Eq.\ (\ref{eq:1129}) does rotate the
whole space around $\boldsymbol{n}$ by an angle $\phi$ while
keeping $\boldsymbol{n}$ invariant. For example, when $\boldsymbol{n}=\boldsymbol{n}_{z}\equiv(0,0,1)^{T}$
we have 
\begin{equation}
R(\boldsymbol{n}_{z},\phi)=\left(\begin{array}{ccc}
c & s & 0\\
-s & c & 0\\
0 & 0 & 1
\end{array}\right),\label{eq:1208-2}
\end{equation}
which is the familiar form of a rotation in the $x-y$ plane around
the $z$ axis.

For each of the six conditions in Eq.\ (\ref{eq:1202-1}) we can get a
bisecting rotation matrix from Eq.\ (\ref{eq:1129}). We use the symbol $R^{(\pm ij)}$
to denote these bisecting rotations: 
\[
R^{(\pm ij)}\equiv R(\boldsymbol{n}|_{n_{i}=\pm n_{j}},\phi)\,.
\]
As an example, for $n_{1}=n_{3}$, we have 
\begin{equation}
R^{(13)}=\left(\begin{array}{ccc}
d & a & p\\
b & h & a\\
q & b & d
\end{array}\right),\label{eq:1202-2}
\end{equation}
where $a=sn_{1}+(1-c)n_{1}n_{2}$, $b=(1-c)n_{1}n_{2}-sn_{1}$, $d=c+n_{1}^{2}(1-c)$. The remaining parameters 
$p$, $q$ and $h$ are determined by $RR^{T}=1$ if $a$, $b$, $d$
are fixed. In general they are not equal to each other, but their precise forms are  
not important here. The point we should notice here is, if $R$ has $n_{1}=n_{3}$,
then the 12-element equals the 23-element, the 21- the 32- and the 11- 
the 33-element. Conversely, if an $SO(3)$ matrix has the form of (\ref{eq:1202-2}),
then it must be a bisecting rotation with its axis on the $x=z$ plane.
This can be seen by solving (\ref{eq:1202-2}) as an equation for $(\boldsymbol{n},\phi)$
(the solution always exists since (\ref{eq:1129}) contains all possible
$SO(3)$ matrices) and finding that the solutions always have $n_{1}=n_{3}$. 

$R$ can be diagonalized by 
\begin{equation}
U_{R}^{\dagger}RU_{R}={\rm diag}(e^{i\phi},1,e^{-i\phi})\,,\label{eq:0109-2}
\end{equation}
where the eigenvalues only depend on $\phi$ while $U_{R}$ only
depends on $\boldsymbol{n}$. As
one can check numerically or by direct analytic calculation, 
$|U_{R}|$ has the following form,
\begin{equation}
|U_{R}|^{2}=\frac{1}{2}\left(\begin{array}{ccc}
1-n_{1}^{2} & 2n_{1}^{2} & 1-n_{1}^{2}\\
1-n_{2}^{2} & 2n_{2}^{2} & 1-n_{2}^{2}\\
1-n_{3}^{2} & 2n_{3}^{2} & 1-n_{3}^{2}
\end{array}\right).\label{eq:1130}
\end{equation}
Here $|U_{R}|^{2}$ is not $|U_{R}| |U_{R}|$ but each element $x_{ij}$ of $|U_{R}|^{2}$  is the absolute value squared of the $ij$-element of $U_R$. 
Note that the order of the columns in (\ref{eq:1130}) can be changed since 
reordering of the columns of a diagonalization matrix is just a matter of permutation of eigenvectors. 
For $n_{1}=n_{3}$, we recommend to write it in this order so that
once one takes $n_{1}=n_{3}$ one immediately obtains a symmetric matrix.
For the other cases such as $n_{1}=n_{2}$ etc., we can always reorder 
the columns to get a symmetric matrix. 

From Eq.\ (\ref{eq:1130}), the proof of Theorem A is easy. One just sets Eq.\ 
(\ref{eq:1130}) equal to its transpose and finds $n_{1}^{2}=n_{3}^{2}$. There are other two
possible permutation of the columns where $(n_{1}^{2}, n_{2}^{2}, n_{3}^{2})^{T}$
is the first or the last column of $U_R$, from which we can get $n_{2}^{2}=n_{3}^{2}$
or $n_{1}^{2}=n_{2}^{2}$. 

This completes our proof of Theorem A.

\section{\label{sec:Generalization}Generalization to the complex case}

The previous theorem only applies for flavor symmetries with real
representations, while some groups used in flavor symmetry model building enjoy 
complex representations. For the complex case, we cannot find a clear geometrical
picture as was possible for real representations in 3-dimensional Euclidean space. 
However we can somewhat generalize the previous theorem to the complex case by finding some connections
between the real and complex case \footnote{Note that the previous real case does not mean there is no CP violation,
while here the complex case does not necessarily mean CP violation.
For example, a bisecting rotation around the axis $(1,1,1)^{T}$ produces a mixing
matrix equal to the Wolfenstein matrix, which has non-zero Jarlskog
invariant. Actually we will see in a 3-dimensional complex representation
of $\Delta(96)$ that the generated mixing scenario has a zero Dirac CP phase.}.  
In the following discussion, all unitary matrices are elements of $SU(3)$
since the difference between $U(3)$ and $SU(3)$ is a trivial phase. \\

{\bf Theorem B:}

{\it If an $SU(3)$ matrix $T$
can be rephased to a real matrix $R$ as follows,
\begin{equation}
T={\rm diag}(e^{i\alpha_{1}},e^{i\alpha_{2}},e^{i\alpha_{3}}) R\, {\rm diag}(e^{i\beta_{1}},e^{i\beta_{2}},e^{i\beta_{3}})\,, \label{eq:1130-2}
\end{equation}
and if the $R$ is one of the bisecting rotations with $n_{i}=n_{j}$ \footnote{ $n_{i}=-n_{j}$ is not necessary to be considered here because the minus sign can always be absorbed by rephasing in the complex case.} from Theorem A, 
and if further 
\begin{equation}
\alpha_{k}+\beta_{k}=0\thinspace(k\neq i,j)\,,\label{eq:1130-1}
\end{equation}
then $T$ gives a symmetric mixing matrix \footnote{Strictly speaking, we should write $\alpha_{k}+\beta_{k}=0=0\,({\rm mod}\,2\pi)$ 
or $\alpha_{k}+\beta_{k}=0=2n\pi$ with any integer $n$. For simplicity,
we abandon $({\rm mod}\,2\pi)$ in all expressions.}.}\\

Note on Theorem B: $k$ in Eq.\ (\ref{eq:1130-1}) is the remaining 
number among $\{1,2,3\}$ when $|n_{i}|=|n_{j}|$ picks out two numbers
for $i$ and $j$. Since the rephasing matrices diag$(e^{i\alpha_{1}},e^{i\alpha_{2}},e^{i\alpha_{3}})$
and diag$(e^{i\beta_{1}},e^{i\beta_{2}},e^{i\beta_{3}})$ should be
in $SU(3)$, it must hold $\alpha_{1}+\alpha_{2}+\alpha_{3}=\beta_{1}+\beta_{2}+\beta_{3}=0$.
So actually Eq.\ (\ref{eq:1130-1}) is equivalent to 
$\alpha_{i}+\alpha_{j}=-\beta_{i}-\beta_{j}$.

As an example, consider that the bisecting rotation is $R^{(13)}$
in Eq.\ (\ref{eq:1202-2}), then we have $\alpha_{2}+\beta_{2}=0$. In this case 
$T$ is 
\begin{equation}
T^{(13)}=\left(\begin{array}{ccc}
f+ig & a\eta_{1} & p\eta_{5}\\
b\eta_{3} & h & a\eta_{2}\\
q\eta_{6} & b\eta_{4} & f-ig
\end{array}\right),\label{eq:1204}
\end{equation}
where $\eta_{i}$ are some phases, i.e.\ $|\eta_{i}|=1$. The 22-element 
is still $h$ the (real) and 11-element is the conjugate of the 33-element, as a result of $\alpha_{2}+\beta_{2}=0$. 
We also have $\eta_{1}\eta_{2}\eta_{3}\eta_{4}=1$ because $\alpha_{1}+\alpha_{3}+\beta_{1}+\beta_{3}=0$. 
$T^{(13)}$ can be diagonalized by a unitary matrix which we call
$U_{T13}$ and one can check that 
\begin{equation}
|U_{T13}|^{2}=\left(\begin{array}{ccc}
\frac{t}{2}+\frac{g}{2s_{\varphi}} & 1-t & \frac{t}{2}-\frac{g}{2s_{\varphi}}\\
1-t & \frac{h-c_{\varphi}}{1-c_{\varphi}} & 1-t\\
\frac{t}{2}-\frac{g}{2s_{\varphi}} & 1-t & \frac{t}{2}+\frac{g}{2s_{\varphi}}
\end{array}\right),\label{eq:1204-1}
\end{equation}
where $c_{\varphi}=\cos\varphi$, $s_{\varphi}=\sin\varphi$ and
\begin{eqnarray}
c_{\varphi} & = & (-1+2f+h)/2\label{eq:1204-3}\,,\\
t & = & 1-\frac{1-h}{2(1-c_{\varphi})}\,.\label{eq:1204-2}
\end{eqnarray}
We can see that indeed $|U_{T13}|=|U_{T13}|^{T}$.

\section{\label{sec:Application}Application}
In this section we will apply our theorems to an actual mixing scheme.  
After the T2K neutrino experiment measured a large non-zero $\theta_{13}$
in 2011 \cite{T2K:2011sj}, many models have been proposed to explain the result. 
Refs.\ \cite{Toorop:2011jn,Toorop:2011re} scanned a
series of discrete groups ($\Delta(6n^{2})$ and $\Gamma_{N}$), and one of the found schemes was quite close to the T2K result at the time. In the standard parametrization, the angles are 
\begin{equation}
\theta_{23}=\theta_{12}=\tan^{-1}\frac{2}{\sqrt{3}+1}=36.21^{\circ}\,,\label{eq:1204-7}
\end{equation}
and 
\begin{equation}
\theta_{13} =\sin^{-1}\left(\frac{1}{2}-\frac{1}{2\sqrt{3}}\right)=12.20^{\circ}\,.\label{eq:1204-4}
\end{equation}
In total the PMNS matrix is 
\begin{equation}
U=\left(\begin{array}{ccc}
\frac{1}{6}\left(3+\sqrt{3}\right) & \frac{1}{\sqrt{3}} & \frac{1}{6}\left(3-\sqrt{3}\right)\\
-\frac{1}{\sqrt{3}} & \frac{1}{\sqrt{3}} & \frac{1}{\sqrt{3}}\\
\frac{1}{6}\left(3-\sqrt{3}\right) & -\frac{1}{\sqrt{3}} & \frac{1}{6}\left(3+\sqrt{3}\right)
\end{array}\right).\label{eq:1204-6}
\end{equation}
While this mixing scheme is ruled out by current data, it fulfills our criterion of a symmetric mixing matrix, and could serve as a starting point or zeroth-order approximation.

The mixing scheme can be produced in $\Delta(96)$ group which can
be defined by three generators $a$, $b$ and $c$ with the following properties \cite{Escobar:2008vc}: 
\begin{eqnarray}
a^{3}=b^{2} & = & (ab)^{2}=c^{4}=1\,,\nonumber \\
caca^{-1} & = & a^{-1}c^{-1}a=bcb^{-1}=b^{-1}cb\,,\nonumber \\
cbc^{-1}b^{-1} & = & bc^{-1}b^{-1}c\,.\label{eq:0109}
\end{eqnarray}
In a 3-dimensional faithful representation, $a$, $b$ and $c$ can be represented
by \cite{Escobar:2008vc}
\begin{eqnarray}
a^{(3)} & = & \left(\begin{array}{ccc}
0 & 1 & 0\\
0 & 0 & 1\\
1 & 0 & 0
\end{array}\right),\,\,b^{(3)}=\left(\begin{array}{ccc}
0 & 0 & -1\\
0 & -1 & 0\\
-1 & 0 & 0
\end{array}\right),\,\,\nonumber \\
c^{(3)} & = & \left(\begin{array}{ccc}
i & 0 & 0\\
0 & -i & 0\\
0 & 0 & 1
\end{array}\right).\label{eq:0109-1}
\end{eqnarray}
The mixing scheme is produced if a $\mathbb{Z}_{3}$ subgroup generated by $T=a^{2}(cb)^{2}c$
and a  $\mathbb{Z}_{2}\otimes\mathbb{Z}_{2}$ subgroup generated by $S_{1}=b,S_{2}=a^{2}c^{2}ac^{2}$ is 
applied to charged leptons and neutrinos respectively. To be precise, 
\begin{equation}
\mathbb{Z}_{3}:T=\left(\begin{array}{ccc}
0 & 0 & i\\
-1 & 0 & 0\\
0 & i & 0
\end{array}\right)\label{eq:1102-6}
\end{equation}
and the two $\mathbb{Z}_{2}$ are generated by  
\begin{equation}
\mathbb{Z}_{2}\otimes\mathbb{Z}_{2}:S_{1}=-\left(\begin{array}{ccc}
0 & 0 & 1\\
0 & 1 & 0\\
1 & 0 & 0
\end{array}\right),\,\,\,\,\,S_{2}=\left(\begin{array}{ccc}
-1 & 0 & 0\\
0 & 1 & 0\\
0 & 0 & -1
\end{array}\right).\label{eq:1102-2-1}
\end{equation}
From our theorem, it is easy to see that this can produce symmetric
mixing. In the diagonal neutrino basis we transform $T$ to $T_{d}$, 
\begin{equation}
T_{d}=\left(\begin{array}{ccc}
\frac{i}{2} & \frac{i}{\sqrt{2}} & \frac{i}{2}\\
-\frac{1}{\sqrt{2}} & 0 & \frac{1}{\sqrt{2}}\\
-\frac{i}{2} & \frac{i}{\sqrt{2}} & -\frac{i}{2}
\end{array}\right)\label{eq:1206}
\end{equation}
and $(S_{1},S_{2})$ to $(S_{1d},S_{2d})$: 
\begin{equation}
S_{1d}={\rm diag}(-1,-1,1)\,;\,\,S_{2d}={\rm diag}(-1,1,-1)\,.\label{eq:1206-1}
\end{equation}
Then, according to our theorem, we see that $T_{d}$ can be rephased via a transformation defined in Eq.\ (\ref{eq:1130-2}) to
a bisecting rotation $R$ with $\alpha_{2}=\beta_{2}=0$, 
$(\alpha_{1},\alpha_{3})=(\pi/2,-\pi/2)$ and $(\beta_{1},\beta_{3})=(0,0)$.
Or, in a simpler way, $T_{d}$ has the form of Eq.\ (\ref{eq:1204}).
So the mixing matrix should be symmetric.\\ 

A dynamical realization of the mixing scheme in $\Delta(96)$ has been studied
in Ref.\ \cite{Ding:2012xx}. That model is rather complicated using 
both 3- and 6-dimensional representations. Here we present
a simpler model which only uses two additional sets of scalar fields 
$\phi^{\nu},\phi^{\ell}$, and features all particles in the same 3-dimensional
representation of $\Delta(96)$ of (\ref{eq:1102-6}) and (\ref{eq:1102-2-1}).
\begin{equation}
\ell,\ell^{c},\nu,\phi^{\nu},\phi^{\ell}\sim3\,.\label{eq:1206-2}
\end{equation}
We use the representation in (\ref{eq:1102-6}) and (\ref{eq:1102-2-1})
rather than (\ref{eq:1206}) and (\ref{eq:1206-1}) because the Clebsch\textendash Gordan (CG)
coefficients are simpler. The result does not depend on the basis. 
The CG coefficients we will use 
in this representation are 
\begin{eqnarray}
3\otimes\overline{3} & \rightarrow & 1:\delta_{ij}\label{eq:1215}\\
3\otimes3\otimes3 & \rightarrow & 1:\epsilon_{ijk}\label{eq:1215-1}\\
3\otimes3\otimes3\otimes3 & \rightarrow & 1:\delta_{ijmn}\label{eq:1215-2}\\
3\otimes3\otimes\overline{3}\otimes\overline{3} & \rightarrow & 1:\delta_{im}\delta_{jn}
\delta_{in}\delta_{jm} \,.\label{eq:1215-3}
\end{eqnarray}
Here $\epsilon_{ijk}$ is the Levi-Civita tensor (or order 3 antisymmetric
tensor) and $\delta_{ijmn}$ is defined as 
\begin{equation}
\delta_{ijmn}=\begin{cases}
1 & (i=j=m=n)\,,\\
0 & \textrm{otherwise.}
\end{cases}\label{eq:1206-3}
\end{equation}
The invariant Lagrangian in the lepton sector is 
\begin{eqnarray}
\mathcal{L} & = & y_{1}^{\ell}\epsilon_{ijk}\phi_{i}^{\ell}\ell_{j}\ell_{k}^{c}+y_{2}^{\ell}\delta_{ijmn}\phi_{i}^{\ell}\phi_{j}^{\ell}\ell_{m}\ell_{n}^{c}+y_{3}^{\ell}\delta_{im}\delta_{jn}\overline{\phi}_{i}^{\ell}\overline{\phi}_{j}^{\ell}\ell_{m}\ell_{n}^{c}\nonumber \\
 &  & +y_{1}^{\nu}\delta_{ijmn}\phi_{i}^{\nu}\phi_{j}^{\nu}\nu_{m}\nu_{n}
+y_{2}^{\nu}\delta_{im}\delta_{jn}\phi_{i}^{\nu}\phi_{j}^{\nu}\nu_{m}\nu_{n}\,.\label{eq:1215-4}
\end{eqnarray}
After symmetry breaking, 
$\phi^{\nu}$ and $\phi^{\ell}$ obtain the following VEVs: 
\begin{equation}
\left\langle \phi^{\ell}\right\rangle =v^{\ell}(1,-1,i) \,,\,\,
\left\langle \phi^{\nu}\right\rangle =v^{\nu}(1,0,1)\,.\label{eq:1208}
\end{equation}
In the charged lepton sector $M_{\ell}=m_{\ell} m_{\ell}^{\dagger}$ is 
\begin{equation}
M_{\ell}=|v^{\ell}|^{2}\left(\begin{array}{ccc}
u & x+iy & -ix-y\\
x-iy & u & ix-y\\
ix-y & -ix-y & u
\end{array}\right),\label{eq:1125-1}
\end{equation}
while the neutrino mass matrix is 
\begin{equation}
M_{\nu}=|v^{\nu}|^{2}\left(\begin{array}{ccc}
A & 0 & B\\
0 & 0 & 0\\
B & 0 & A
\end{array}\right),\label{eq:1125}
\end{equation}
where $A=y_{1}^{\nu}+y_{2}^{\nu};B=y_{2}^{\nu}$ with 
\begin{equation}
u=2|y_{1}^{\ell}|^{2}+2|y_{3}^{\ell}|^{2}+|y_{2}^{\ell}+y_{3}^{\ell}|^{2}\,\label{eq:1215-5}
\end{equation}
\begin{equation}
x=|y_{1}^{\ell}|^{2}-3|y_{3}^{\ell}|^{2}-2 {\rm Re}[y_{2}^{\ell*}y_{3}^{\ell}]\,,\label{eq:1215-6}
\end{equation}
\begin{equation}
y=2 {\rm Re}[y_{1}^{\ell*}y_{2}^{\ell}]\,.\label{eq:1215-7}
\end{equation}
$M_{\ell}$ and $M_{\nu}$ can be diagonalized by the following unitary
matrices
\begin{equation}
U_{\nu}=\left(\begin{array}{ccc}
\frac{1}{\sqrt{2}} & 0 & -\frac{1}{\sqrt{2}}\\
0 & 1 & 0\\
\frac{1}{\sqrt{2}} & 0 & \frac{1}{\sqrt{2}}
\end{array}\right),\label{eq:1123}
\end{equation}
\begin{equation}
U_{\ell}=\frac{1}{\sqrt{3}}\left(\begin{array}{ccc}
\omega^{2} & 1 & \omega\\
-\omega & -1 & -\omega^{2}\\
-i & -i & -i
\end{array}\right).\label{eq:1123-1}
\end{equation}
Thus, the PMNS matrix is 
\begin{equation}
U_{\rm PMNS}=\left(\begin{array}{ccc}
\frac{i+\omega}{\sqrt{6}} & -\frac{\omega^{2}}{\sqrt{3}} & -\frac{-i+\omega}{\sqrt{6}}\\
\frac{1+i}{\sqrt{6}} & -\frac{1}{\sqrt{3}} & -\frac{1-i}{\sqrt{6}}\\
\frac{i+\omega^{2}}{\sqrt{6}} & -\frac{\omega}{\sqrt{3}} & -\frac{-i+\omega^{2}}{\sqrt{6}}
\end{array}\right).\label{eq:1101-4}
\end{equation}
It is related to the matrix in Eq.\ (\ref{eq:1204-6}) via
\begin{equation}
U_{\rm PMNS}={\rm diag}(e^{i\beta_{1}},e^{i\beta_{2}},e^{i\beta_{3}}) U {\rm diag} (1,e^{i\alpha_{1}},e^{i\alpha_{2}})\,,\label{eq:1215-9}
\end{equation}
where 
$\beta_{1}  =  105^{\circ}$,$\beta_{2}=225^{\circ}$, $\beta_{3}=165^{\circ}$, $\alpha_{1}=45^{\circ}$, $\alpha_{2}=90^{\circ}$. Here $\alpha_{1}$, $\alpha_{2}$ would 
be the Majorana phases if the couplings $y_{1}^{\nu}$ and  $y_{2}^{\nu}$ 
in Eq.\ (\ref{eq:1215-4}) were real. 
Therefore even though the Dirac-type CP is conserved in this model, generally there is still 
CP violation due to the non-zero Majorana phases, unless the 
phases of $y_{1}^{\nu}$ and  $y_{2}^{\nu}$ are tuned to exactly cancel $\alpha_{1}$, $\alpha_{2}$.\\ 

Our theorem can also be applied to the quark sector. One just assigns
the bisecting rotational symmetry to the residual symmetry of up-type
quarks (or down-type) and the mirror symmetries to that of down-type
quarks (or up-type), then the CKM mixing will be symmetric. However
building a realistic model for the CKM mixing is a somewhat more difficult task.
Compared to the lepton sector where hundreds of flavor symmetry models
have been proposed, for the quark sector much fewer models exist. This is due to the fact 
that the small CKM mixing angles do not have straightforward geometric interpretation, 
which is the basis of discrete flavor symmetry building. 
Among the existing models for the CKM mixing, we cannot find one that fulfills our 
criteria (exceptions are of course the trivial cases in which one interprets the CKM matrix 
as the unit matrix or as a matrix which only consists of the Cabibbo angle), and 
scanning all discrete groups for the flavor symmetry of quarks is out of the main
purpose of this paper. Anyway, when looking for flavor groups to build
models for the quark sector, our theorem could be a guidance because when a
mixing scheme generated from a flavor symmetry is close to realistic
CKM mixing, then it must be also close to a symmetric form. 
%, no matter whether the symmetric mixing is a starting point or not.

\section{Conclusion}
A possible zeroth-order, but surely aesthetically attractive, mixing Ansatz for the CKM and PMNS matrices is that they are symmetric. The origin of symmetric PMNS and CKM matrices from the viewpoint of flavor symmetry models has been the focus of our paper.  

We have proposed a theorem on the relation between symmetric mixing matrices 
and geometric properties of discrete flavor symmetry groups. An illustrative connection between the rotation axes of the geometric body associated to the symmetry group exists, and shows that popular subgroups of $SO(3)$ such as $A_4$ and $S_4$ can lead to symmetric mixing matrices.

Groups with complex irreducible representations do not easily allow for a geometrical interpretation, but a partial generalization of our theorem is possible, which can then apply to $SU(3)$ subgroups such as $\Delta(96)$. A previously studied mixing scheme that turns out to correspond to a symmetric PMNS matrix was used as an explicit example. 

The connection of geometric properties of discrete groups and possible features of the mixing matrices may have further applications.

\begin{acknowledgments}
We thank Yusuke Shimizu for useful discussions. WR is supported by
the Max Planck Society in the project MANITOP. XJX is supported by
the China Scholarship Council (CSC).
\end{acknowledgments}

\bibliographystyle{apsrev4-1}
\bibliography{ref}

\end{document}